\documentclass[twocolumn,twoside,slac_two]{revtex4}
\usepackage{graphicx}
\usepackage{fancyhdr}
\usepackage{xspace}

\newcommand{\Figref}[1]{Figure \ref{fig:#1}}
\newcommand{\ie}[0]{{i.e.}\xspace}
\newcommand{\eg}[0]{{e.g.}\xspace}

\begin{document}

\title{Using TMine for the Fermi-LAT Event Analysis}

\author{Alex Drlica-Wagner\footnote{kadrlica@stanford.edu} and Eric Charles\footnote{echarles@slac.stanford.edu} for the Fermi-LAT Collaboration}

\affiliation{W. W. Hansen Experimental Physics Laboratory, Kavli Institute for Particle Astrophysics and Cosmology, Department of Physics and SLAC National Accelerator Laboratory, Stanford University, Stanford, CA 94305, USA}

\begin{abstract}
The Large Area Telescope (LAT) event analysis is the final stage in the event reconstruction responsible for the creation of high-level variables (\eg, event energy, incident direction, particle type, etc.).  We discuss the development of TMine, a powerful new tool for designing and implementing event classification analyses (\eg, distinguishing photons from charged particles).  TMine is structured on ROOT, a data analysis framework that is the de-facto standard for current high energy physics experiments; thus, TMine fits naturally into the ROOT-based data processing pipeline of the LAT.  TMine provides a visual development environment for the LAT event analysis and utilizes advanced multivariate classification algorithms implemented in ROOT.  We discuss the application of TMine to the next iteration of the event analysis (Pass 8), the LAT charged-particle analyses, and the classification of unassociated LAT $\gamma$-ray sources.
\end{abstract}

\maketitle

\thispagestyle{fancy}

\section{Fermi-LAT Event Analysis} \label{eventanalysis}
The Large Area Telescope (LAT) operates in a low Earth orbit, where every second thousands of particles trigger the detector.  After on-board filtering, the recorded data from these triggers are transmitted to the ground and undergo full event reconstruction.  The final stage of LAT reconstruction is the event analysis, which combines information from each detector subsystem (the anticoincidence detector, tracker, and calorimeter) to create a picture of the event as a whole.  From the event picture, high-level science variables (\ie, event energy and incident direction) are assigned. The event analysis must also address the challenging task of separating the desired $\gamma$-ray signal events from charged particle backgrounds~\cite{atwood}.   

The assignment of fundamental quantities such as particle type, energy, and direction is a complex problem, since the LAT accepts particles over a wide range in parameter space (both in energy and incident angle) and event topology (close to detector edges and gaps).  In addition, discrimination against background at a level of 1 part in $10^{6}$ is required to fulfill the LAT science goals. Classic cut-based analyses lack sufficient accuracy and signal efficiency to meet these goals. To achieve the required instrument performance, the LAT event analysis applies classic cuts followed by multivariate classification trees~\cite{chapter}.

Classification trees (and decision trees in general) belong to the larger family of data mining and machine learning algorithms~\cite{cart}. Classification in the context of machine learning focuses on associating an observation to a sub-population based on the traits present in a set of training observations (where the true sup-population is known).  Training of classification trees is performed through binary recursive partitioning, an algorithm that develops a set of logical cuts by iteratively splitting the training data to maximize the separation of the true sub-populations. For the LAT event analysis, the training of classification trees is performed on sets of $\gamma$-ray and cosmic-ray events generated from a full detector Monte Carlo simulation. These logical cuts are trained on variables describing the physical character of an event shower (\eg, the transverse shower size in the calorimeter, the number of excess tracker clusters surrounding the primary particle track, etc.), while the output is a classification of the event (\eg, the type of particle, the quality of direction reconstruction, etc.).

We introduce TMine, a new tool for implementing both cut-based and multivariate classification algorithms. The goal of TMine is to enhance the performance of the event level analysis to improve the LAT instrument response functions (\ie, effective area, energy resolution, and point spread function). Additionally, TMine has been used for studying LAT charged particle events (electrons, positrons, and protons) and the classifying unassociated LAT $\gamma$-ray sources.

%\begin{turnpage}
\begin{figure*}[t]
\centering
\includegraphics[width=\textwidth]{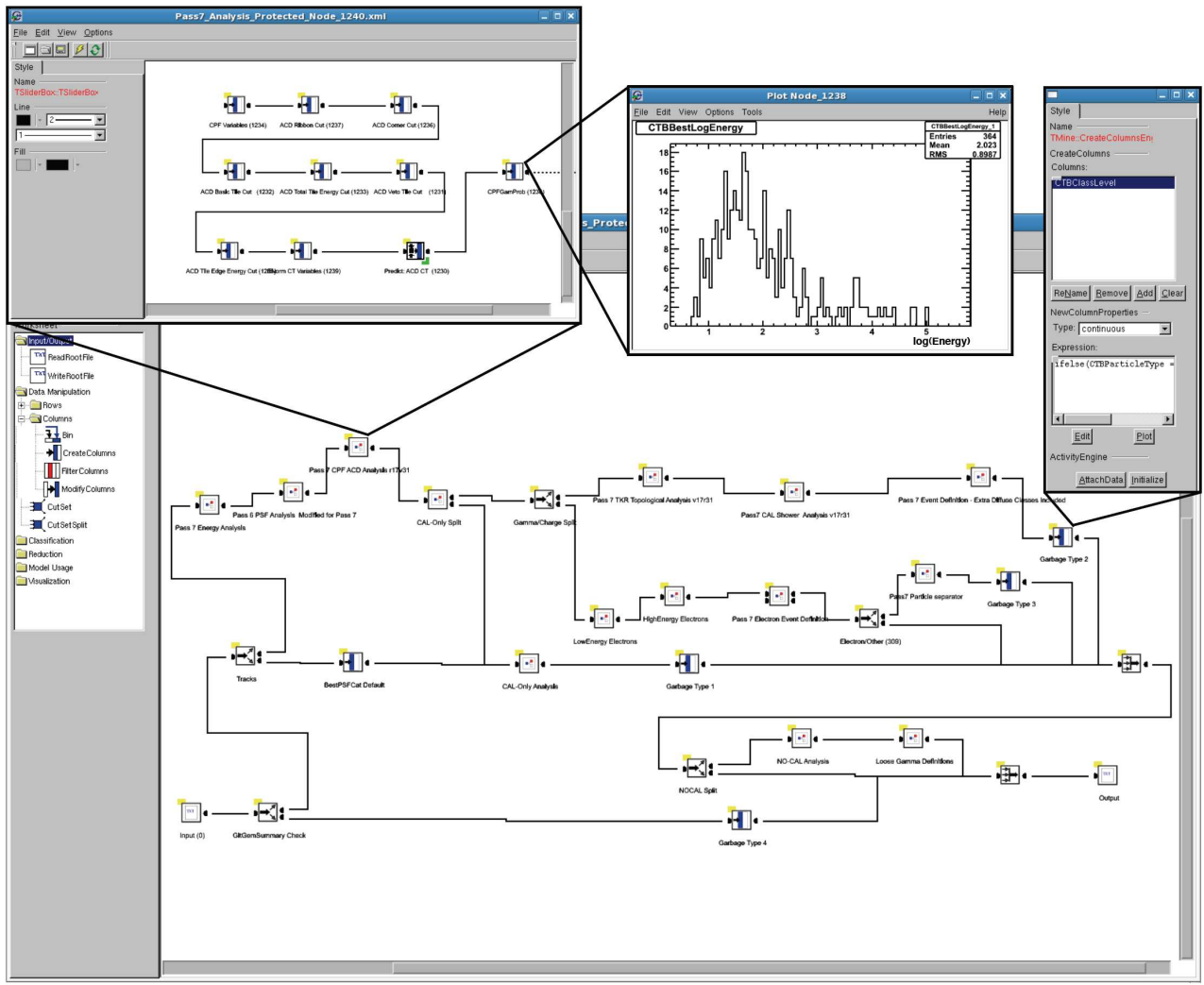}
\caption{The current iteration of the Fermi-LAT event analysis as viewed by TMine.  Many nodes contain sub-analyses (inset top left), while the functionality of each node can be plotted (inset top middle) and edited through a GUI editor (inset top right). A TMine analysis can combine classical cut-based selections with multivariate classification.} \label{fig:worksheet}
\end{figure*}
%\end{turnpage}

\section{The TMine Analysis Tool} \label{tmine}
TMine is an interactive software tool for developing and processing complex event classification analyses.  TMine is based on ROOT~\cite{brun}, the de-facto data analysis framework for current high energy physics experiments. In particular, TMine uses the data set indexing and linking functionality of ROOT to associate newly calculated variables with pre-existing quantities and keeps only the minimal information necessary to process the analysis. Thus, TMine handles large data sets in a quick and efficient manner, especially when some variables are only defined for a small subset of the events.

TMine applies classic event-selection cuts in the standard ROOT manner through TFormulas, TCuts, and event indexing. For the processing and parallel evaluation of sophisticated multivariate classification algorithms, TMine utilizes the ROOT Toolkit for Multivariate Analysis (TMVA)~\cite{hoecker}. Through TMVA, TMine has access to many multivariate classification algorithms including, but not limited to, boosted decision trees and artificial neural networks. While the command-line functionality of ROOT is preserved, the graphical user interface of TMine allows the user to harness the power of ROOT and TMVA in a visual work environment. TMine was specifically designed to address problems faced in high energy physics, though it need not be restricted to these.

A TMine analysis consists of a network of directionally linked nodes controlling work flow and operation (\Figref{worksheet}). Nodes both alter event characteristics (\ie, variable definition, assignment, and selection) and direct events through the network.  Specialized nodes are used for training, testing, and implementing TMVA classification algorithms. Using the machinery of ROOT, TMine is able to split, manipulate, and recombine large quantities of data without excessive duplication of information. Structuring the event analysis in a visual manner has been found to be conceptually powerful when designing the LAT event analysis~\cite{atwood}.

\section{Applications of TMine} \label{applications}
\subsection{The Pass-8 Reconstruction Effort}
\begin{figure}[t]
\centering
\includegraphics[width=\columnwidth]{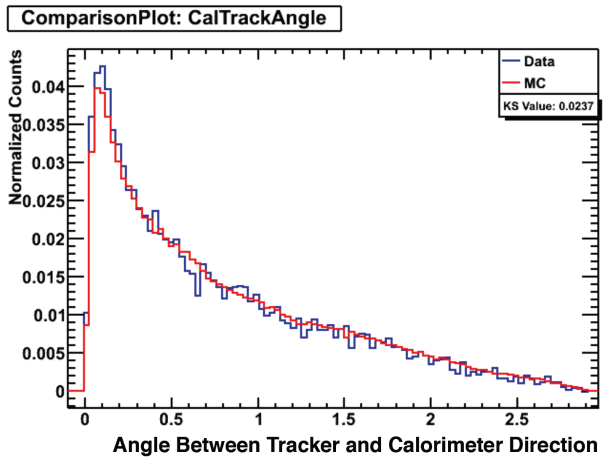}
\caption{Comparisons between a statistical sample of photons from flight data (blue) and simulations (red).  Only variables with good agreement should be used for classification.} \label{fig:datamc}
\end{figure}

Our primary application of TMine is in the development and implementation of the Pass-8 event analysis. The Pass-8 effort is a complete reworking of the LAT simulation and reconstruction software, benefiting from the analysis of flight data (which was unavailable before launch). TMine will improve the interface between event reconstruction and event classification. It also provides improvements to the structure and validation of the Pass-8 event analysis. TMine has built-in functionality for comparing real and simulated data (\Figref{datamc}), an essential step prior to training multivariate classification algorithms~\cite{abdo,ackermann}. Additionally, the TMine interface to TMVA allows for the training of multivariate classification algorithms using larger data sets than was possible with the software tools previously used by the Fermi-LAT Collaboration.

\subsection{LAT Charged-Particle Analyses}
In addition to the Pass-8 effort, TMine has been utilized in a variety of ongoing LAT analyses. Since electromagnetic showers are common to photon, electron, and positron events, the LAT is naturally sensitive to cosmic-ray electrons and positrons~\cite{abdo,ackermann}. For the majority of LAT analyses, these charged particles present a background for $\gamma$-ray science. Thus, the detection of electrons and positrons requires a non-standard event analysis and a reprocessing of the LAT data (the analysis of electrons and positrons has subsequently been appended to the standard event analysis). TMine was used for this reprocessing because it is a stand-alone program that is free from the overhead of the full LAT reconstruction software.

\begin{figure}[t]
\centering
\includegraphics[width=\columnwidth]{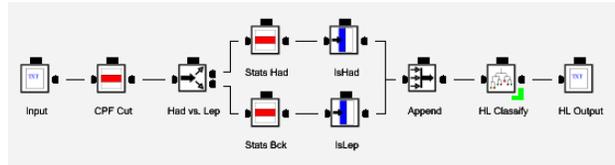}
\caption{A simple TMine worksheet for discriminating cosmic-ray hadron events from cosmic-ray lepton events. Data is input on the left, has a classical charged particle cut applied, and is used to train a TMVA classifier.} \label{fig:proton}
\end{figure}

\begin{figure}[h]
\centering
\includegraphics[width=\columnwidth]{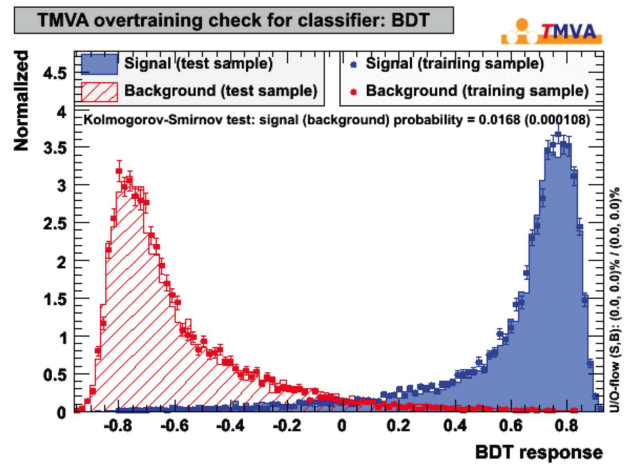}
\caption{Classifier output from the TMine implementation of a TMVA boosted decision tree. Simulated hadrons (marked signal) are distinguished from simulated electrons and positrons (marked background). Events that are hadron-like are assigned positive predictor values, while events that are lepton-like are assigned negative values. The two event classes are well separated, and an independent sample of test events (filled histograms) agrees with the distribution of events used to train the classifier (data points). } \label{fig:tmva}
\end{figure}

A similar effort is underway to study cosmic-ray proton events in more detail~\cite{monzani}. For this task, TMine was used both to design a proton event classification and to reprocess LAT data. The analysis of proton events presents an excellent example of how TMine can be used for event classification. \Figref{proton} shows a simple event analysis for distinguishing hadrons from leptons. This worksheet is read from left to right, with the training data set input on the left and the predicted particle type output on the right. A classic cut selecting charged particles is applied first, followed by a split and tagging of the true particle type. Events are then recombined and used to train TMVA boosted decision trees. The preliminary performance of this classifier when discriminating simulated hadrons from simulated electrons and positrons is shown in~\Figref{tmva}.

\subsection{Classifying Unassociated LAT Sources}
While TMine was originally developed for use with the LAT event analysis, it is not limited to that purpose. Notably, TMine has been utilized to classify unassociated $\gamma$-ray sources~\cite{unassoc}.  Of the 1451 $\gamma$-ray sources in the First LAT Source Catalog (1FGL)~\cite{1FGL}, 630 are unassociated with counterparts in other wavelengths. In an attempt to classify these sources, TMine was used to input individual source characteristics, such as spectral index, spectral curvature, and fractional variability into a forest of TMVA boosted decision trees. These input variables were selected to be independent of source flux, location, or significance, since these distributions differ between associated and unassociated sources. The TMVA decision trees were trained on the set of 1FGL sources already associated with active galactic nuclei (AGN) and pulsars. The output of this analysis was a predictor representing the probability that a source is an AGN versus a pulsar. 

\begin{figure}[t]
\centering
\includegraphics[width=\columnwidth]{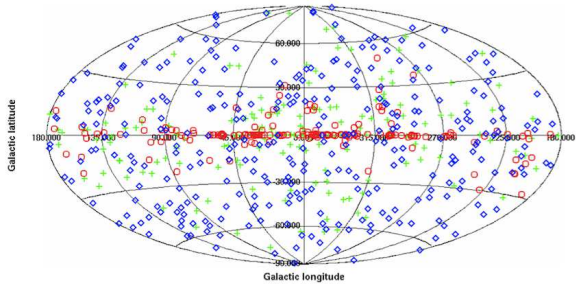}
\caption{Spatial distribution, in Galactic coordinates, for 1FGL unassociated sources classified as AGN candidates (blue diamonds) and pulsar candidates (red circles). As expected, pulsar candidates are distributed primarily along the Galactic plane, while AGN candidates are distributed isotropically. The sources left unclassified are shown as green crosses.} \label{fig:unassoc}
\end{figure}

Unassociated sources were separated into AGN candidates and pulsar candidates by cutting on the output of the classifier. This cut was designed to have $80\%$ efficiency when applied to an independent set of sources associated to AGN and pulsars in the 1FGL. The Galactic latitude of the unassociated sources was explicitly omitted from the classifier training, but the spatial distribution of candidate AGN was found to be isotropically distributed, while the pulsar candidates were distributed along the Galactic plane (\Figref{unassoc}). From follow-up observations on a subset of the unassociated sources, the cut placed on the multivariate classifier is found to be $\sim 70\%$ efficient with a contamination of $\sim5\%$ for both AGN and pulsars~\cite{unassoc}.

\section{Conclusions}
We present TMine, a new tool for developing and processing complex classification tasks. TMine is a ROOT-based tool utilizing the multivariate classification package, TMVA. While the primary application of TMine is to the LAT event analyses (specifically the Pass-8 iteration), it has a wide range of possible applications.

\bigskip
\begin{acknowledgments}
The $Fermi$ LAT Collaboration acknowledges support from a number of agencies and institutes for both development and the operation of the LAT as well as scientific data analysis. These include NASA and DOE in the United States, CEA/Irfu and IN2P3/CNRS in France, ASI and INFN in Italy, MEXT, KEK, and JAXA in Japan, and the K.~A.~Wallenberg Foundation, the Swedish Research Council and the National Space Board in Sweden. Additional support from INAF in Italy and CNES in France for science analysis during the operations phase is also gratefully acknowledged.

ADW is supported in part by the Department of Energy Office of Science Graduate Fellowship Program (DOE SCGF), made possible in part by  the American Recovery and Reinvestment Act of 2009, administered by ORISE-ORAU under contract no. DE-AC05-06OR23100.
\end{acknowledgments}

%\bibliography{bib}

\begin{thebibliography}{}
\bibitem{atwood} W.B. Atwood, et al. Fermi-LAT Collaboration. Ap. J. {\bf 697} 1071-1102 (2009)
\bibitem{chapter} W.B. Atwood, in {\it Advances in Machine Learning and Data Mining for Astrononmy} (Chapman and Hall, London, in press)
\bibitem{cart} L. Breiman, et al. {\it Classification and Regression Trees} (Wadsworth and Brooks, Monteray, 1984)
\bibitem{brun} R. Brun and F. Rademakers, Nucl. Inst. \& Meth. in Phys. Res. A {\bf 389}  81-86 (1997)
\bibitem{hoecker} A. Hoecker, et al. PoS ACAT {\bf 040} (2007)
\bibitem{abdo} A. Abdo, et al. Fermi-LAT Collaboration. Phys. Rev. Lett. {\bf 102} 181101 (2009)
\bibitem{ackermann} A. Ackermann, et al. Fermi-LAT Collaboration. Phys. Rev. D {\bf 82} 092004 (2010)
\bibitem{monzani} M.E. Monzani, American Astronomical Association, HEAD Meeting {\bf 11} 36.04 (2010)
\bibitem{unassoc} M. Ackermann, et al. Fermi-LAT Collaboration. Submitted to Ap.J., arXiv:1108.1202
\bibitem{1FGL} A. Abdo, et al. Fermi-LAT Collaboration. Ap. J. Supp. {\bf 188} 405-436 (2010)
\end{thebibliography}

\end{document}